\documentstyle[12pt]{article}

\def\ot{\otimes}
\def\R{{\cal R}}
\def\C{{\bf C}}
\def\T{{\cal T}}
\def\G{{\cal G}}
\def\B{{\cal B}}
\def\A{{\cal A}}

\def\e{\varepsilon}
\def\tr{\triangleright}
\def\trl{\triangleleft}

\def\jp{{1\over 2}}

\def\la{\langle}
\def\ra{\rangle}

\def\fq{{\rm Fun}_q(SU(2))}
\def\uq{U_q( su(2))}
\def\fp{{\rm Fun}_q(S^2)}

\def\d{\partial_z}
\def\dz{\partial_{\bar z}}
\def\bz{\bar z}
\begin{document}
\begin{flushright}
{}~
IML 2003-12\\
math-ph/0310062
\end{flushright}

\vspace{3pc}
\begin{center}
{\Huge \bf $q$-deformation of $z\to {\alpha z+\beta\over \gamma z+\delta }$}\\ 
[45pt]{\small
{\bf Ctirad Klim\v{c}\'{\i}k}
\\ ~~\\Institute de math\'ematiques de Luminy,
 \\163, Avenue de Luminy, 13288 Marseille, France} 
\end{center}

\vspace{2.2 cm}
\centerline{\bf Abstract}
\vspace{0.2 cm}
We construct the action of the quantum double of $\uq$ 
 on the standard Podle\'s sphere  and interpret it as the quantum projective 
formula
generalizing to the $q$-deformed setting the 
action of the
Lorentz group of global conformal transformations 
on the
ordinary Riemann sphere.  
\newpage 
\section{Introduction}
As it is well-known, the Lorentz group 
$SL(2,\C)$ naturally acts
on the Riemann sphere $S^2$ by the
conformal transformations
$$z\to {\alpha z+\beta\over \gamma z+\delta }, \quad \alpha,\beta,\gamma,\delta\in {\C} ,
\quad \alpha\delta-\beta \gamma=1.\eqno(1)$$
The $q$-deformation of  $S^2$ is referred to as
 the [Podle\'s] sphere.
 One of the goals of this note
 is to find the corresponding  $q$-deformation
 of the
conformal transformations (1).  We note that 
the restriction 
$\bar \alpha=\delta, \bar \gamma=-\beta$ gives the action of the 
group $SU(2)$ on $S^2$
which just corresponds to the standard geometrical
 rotations of the 
two-sphere embedded into three-dimensional 
Euclidean space. The q-deformed
version of this $SU(2)$ action was studied  in detail 
by [Podle\'s].
   However,
to our best knowledge, the $q$-conformal 
action of the full $q$-Lorentz group on the 
Podle\'s sphere has not yet been
reported.

Recall, that the algebra of functions Fun$(S^2)$ 
on ordinary
 two-sphere 
can be viewed as the algebra of those functions 
on the group $SU(2)$
which are invariant with respect to the right
 action of the    maximal  torus
$U(1)$ on   $SU(2)$. This gives  the dual   
description of the coset
$SU(2)/U(1)\equiv S^2$. The group $SU(2)$ acts 
naturally
from the left on this coset and this action can be  
extended to the action (1) of the Lorentz group 
$SL(2,\C)$ on $S^2$ since we have a well-known 
identification
$SU(2)/U(1)\equiv SL(2,{\C})/B$ with $B$ being 
the Borel subgroup
(consisting of uppertriangular matrices) of
 $SL(2,\C)$.

In order to construct the $q$-deformation
 of the picture just described,
we take some inspiration from the theory 
of Poisson-Lie groups (see
   [Semenov-Tian-Shansky, Klim\v c\'{\i}k]
 for the elements). 
There is the Iwasawa decomposition 
$SL(2,{\C})=SU(2)AN$ of the Lorentz
group where $AN$  is the subgroup of 
uppertriangular complex
 $2\times 2$ matrices
with real positive numbers on diagonal and
 unit determinant. 
$AN$ turns out to be the dual Poisson Lie
 group of $SU(2)$ and
$SL(2,{\C})$ is the Drinfeld double of 
$SU(2)$ in the Poisson-Lie
sense of this word.
Now the Drinfeld double  $SL(2,\C)$ acts 
on Fun($SU(2)$): the action
of its subgroup 
$SU(2)$ is induced just by the left
 multiplication of $SU(2)$
on itself and the subgroup
$AN$ acts by the so-called dressing 
tranformations. This action
of the $SL(2,\C)$ on Fun($SU(2)$) descends 
to Fun($SU(2)/U(1)(\equiv S^2)$ and it turns out 
to be given by the projective action (1), where $z$ is the
standard complex coordinate on the Riemann sphere.

The theory of Poisson-Lie groups is a sort
 of the semiclassical
limit of the theory of $q$-deformed Hopf 
algebras for the
deformation parameter $q$ approaching $1$. 
Many Poisson-Lie
concepts can be directly generalized to 
the Hopf algebra setting
like e.g. the Drinfeld double or the 
dressing transformations.
In particular, the Poisson-Lie concept of 
the duality  translates
into the famous [Drinfeld] duality in the 
world of Hopf algebras.
Having in mind the parallels between the 
Poisson-Lie and  the Hopf worlds, it is not 
difficult to find 
the $q$-deformation
of the projective formula. We proceed as 
follows.

   The Podle\'s sphere $\fp$
is a one-parameter  deformation 
of the algebra  Fun$(S^2)$. It is generated
 by the $U(1)$ right-invariant 
elements  of the quantum group $\fq$. 
The deformed envelopping algebra $\uq$   
naturally
acts on the deformed $\fq$. (This 
corresponds to the  left action
of  $SU(2)$ on  Fun($SU(2)$) just described
 above).  The Hopf dual
$U_q(an)$ of $\uq$ acts on $\fq$ in the
 Hopf-dressing way.
(This corresponds to the dressing action 
of $AN$ on  Fun($SU(2)$).
As noted by [Korogodsky], the Hopf analogue
 of the
dressing action of $U_q(an)$ on $\fq$
is the adjoint action of the Hopf algebra 
$\fq$ on itself. This
statement
is consistent due to the Drinfeld duality  
 isomorphism between $U_q(an)$ and $\fq$.
The respective actions of $U_q(an)$ and 
$\uq$ on $\fq$ combine to the
action of the Drinfeld double  $D(\uq)$  
on $\fq$. This Drinfeld double is nothing 
but the $q$-Lorentz group
(see [Podle\'s $\&$ Woronowicz])  and the 
only consistency
check of the construction consists in 
verifying that 
the action of the $q$-Lorentz group 
descends from $\fq$ on $\fp$.
It turns out to be  the case and thus  we obtain the
 $q$-deformation of the 
projective formula [1].

In section 2, we describe the action of 
the quantum 
double  $D(U_q(su(2)))$ on the Podle\'s sphere and in section 3 we
  show that it leads  to the projective formula (1) in the 
limit $q\to 1$. We finish with a short outlook.
 
\section{Action of the Drinfeld  double $D(\uq)$ on
 the Podle\' s sphere}

First we  recall some relevant facts concerning (the $*$-actions 
 of) the Drinfeld double.  The reader can mostly find them also  in
 [Majid], however, our exposition between Eqs. (9) and  (11)
 is original.

 Thus let $H$ be a Hopf algebra, 
$\tilde H$   its dual and $H^{cop}$ the coopposite Hopf algebra of $H$.
The Drinfeld double $D(H)$  is another Hopf 
algebra which is generated by its two sub-Hopf algebras
$H^{cop}$ and $\tilde H$.
The  coalgebra structure of $D(H)$  is  just  that of 
$H^{cop}\ot \tilde H$, the antipode $S_D$ is given by
 $$S_D(U\ot  f)\equiv (1\ot Sf)(S^{cop}U\ot 1)=$$
 $$=(S^{cop}U)''\ot (Sf)''<(S^{cop}U)',(Sf)'><S^{cop}(S^{cop}U)''',(Sf)''' >\eqno(2)$$
 and  the product is defined by the
following cross relations  [Majid]: 
$$\la U',f'\ra (U''\ot 1)(1\ot f'')=\la U'',f''\ra (1\ot f')(U'\ot 1).
\eqno(3)$$
Here $U\in H^{cop}$, $f\in \tilde H$,  $\la .,.\ra$
 is the duality pairing between $H$ and $\tilde H$ 
and we use the Sweedler notation for the
 coproduct
$$\Delta^{cop}(U)=\sum_p U'_p\ot U''_p\equiv
 U'\ot U'', \quad \Delta(f)=f'\ot f''.\eqno(4)$$
The formula (3) is particularly useful if 
we know the generators and 
their relations for both algebras $H^{cop}$ and 
$\tilde H$ separately.
The set of relations for the algebra structure
of  $D(H)$ can be then directly obtained 
from (3) and (4).  
 
 If, moreover,  $H$ and $\tilde H$ are equipped with a compatible star structures, then  the quantum double $D(H)$ can be also naturally made a $*$-Hopf algebra.
 Recall that  a star $*$ on $H$ is a antilinear  antihomorphism of $H$ satisfying 
 $S*S*=Id$, $*^2=Id$,$(*\otimes *)\Delta =\Delta *$ and 
 $ *\e =\e *$. The standard compatibility relation (cf.[Majid]) between the 
  stars on $H$ and $\tilde H$ reads
  $$<U^*,f>=\overline{<U, (Sf)^*>}, \quad U\in H, f\in \tilde H.\eqno(5)$$
The explicit formula for the star $*$  on $D(H)$ is  then uniquely determined as follows
  $$(U\ot f)^*\equiv (1\ot f^*)(U^*\ot 1)=$$
  $$=U^{*''}\ot f^{*''}<U^{*'}, f^{*'}><S^{cop}U^{*'''},f^{*'''}>, \quad U\in H^{cop}, f\in\tilde H,\eqno(6)$$
  where the star on $H^{cop}$ is the same as that on $H$.
  
The algebras $H^{cop}$ and $\tilde H$    act   
(from the left) on $\tilde H$ 
respectively as 
$$U\triangleright h= \la S^{cop}(U),h'\ra h'',
\quad U\in H^{cop} , h\in \tilde H,
\eqno(7a)$$
$$f\triangleright h= f'hS(f''),\quad f,h \in \tilde H .\eqno(7b)$$
  We note that $S^{cop}=S^{-1}$, where $S$ is the antipode of $H$.
Using the basic axioms of Hopf algebras, it
is easy to  check that the definitions ($7ab$)
 imply
$$\la U',f'\ra U''\tr (f''\tr h)=\la U'',f''
\ra f'\tr (U'\tr h). 
   \quad U\in H^{cop}, \quad f,h\in \tilde H.$$
By comparing with the defining relation (3), 
this means 
 that ($7ab$) describes in fact the left 
action of the 
quantum double $D(H)$ on $\tilde H$. Explicitely:
  $$(U\ot f)\tr h \equiv U\tr(f\tr h),  \quad U\in H^{cop}, \quad f,h\in \tilde H\eqno(8)$$
   It can be also
 directly checked (with the help of the condition (5)), that
this action is compatible with the algebra structure of  $\tilde H$  
    and with the $*$-structure on $\tilde H$. Explicitely:
   $$x\tr (fh)=  (x'\tr f)(x''\tr h), \quad x\in D(H), f,h \in \tilde H,\eqno(9a) $$ 
     $$(x\tr f)^*=(S_D(x))^*\tr f^*, \quad x\in D(H), f\in \tilde H. \eqno(9b)$$

  \vskip1pc
 
  \noindent Now let $k\in H^{cop}$ be a group-like  selfadjoint element, i.e. $k^*=k$,
  $ \Delta^{cop}k=k\ot k$, $\e(k)=1$.  We can then define a linear space $A$ consisting
   of invariant
  elements of $\tilde H$ with respect to the right  action of $k$ and $S(k)$ on $\tilde H$ :
  $$A=\{f\in \tilde H, \quad  <f'',k>f'\equiv f\trl k =f, \quad f\trl S(k)=f\}.\eqno(10)$$
  We have for $f,g\in A$
  $$(fg)\trl k= <f''g'',k>f'g'=<f'',k><g'',k>f'g'=fg$$
  and, in the same way, $(fg)\trl S(k)=1$
  which means that $A$ is the subalgebra of $\tilde H$.   We obtain easily also
  the $*$-stability of $A$, since for $f\in  A$ we have
  $$f^*\trl k= <f^{*''},k>f^{*'}=(<f'',(S(k))^*>f')^*=(f\trl S(k))^*=f^*$$
  and, in the same  way, $f^*\trl S(k)=(f\trl k)^*=f^*.$
  
  It is not difficult to prove that
  $A$ is also stable with respect to the action ($7ab$) of the quantum double $D(H)$ on $\tilde H$.
  Indeed,  we have for the $H^{cop}$ action ($7a$):
  $$ U\tr(f\trl k)=<S^{cop}(U),f'>f''<f''',k>= (U\tr f)\trl k, \quad U\in H^{cop}, f\in \tilde H.$$
  The proof of stability for $\tilde H$ action ($7b$) is slightly more involved:
  $$(h\tr f) \trl k =<(h'fS(h''))'',k>((h'fS(h''))'=<h''f''Sh''',k>h'f'Sh''''=$$
$$=  <h'',k><f'',k><Sh''',k>h'f'Sh''''=<h'', kS(k)><f'',k>h'f'Sh'''=$$ 
$$= <f'',k>h'f'Sh''=
 h\tr(f\trl k), \quad h,f\in \tilde H.$$
 The same formulae hold true upon replacing $k\to S(k)$.

  \vskip1pc

\noindent In the context of our paper, the $*$-Hopf algebra  $H$    will be
the  standard   deformation  $\uq$ of $U(su(2))$, 
$\tilde H$ will be the corresponding
dual deformation 
$\fq$ of Fun($SU(2)$) and $A$ will be  the Podle\'s sphere $\fp$.
  For the sake of mathematical rigour, we should pay attention to the fact that  the notion
  of the dual Hopf algebra needs some clarification in the
  infinite-dimensional case.  Actually,  $\uq$ and $\fq$ are in duality
  in the sense of chapters V.7 and VII.4  of the book of [Kassel].
   The general Drinfeld double formulae (2) - (10)  then work with this notion of duality
  with the bilinear pairing given by Eq. (12d).

For  the description  of the Hopf algebras  $\fq$ and $\uq$,
we use  the conventions of   [D\c abrowski $\&$ Sitarz] and [D\c abrowski et al]. 
 Thus let $q\neq 1$ be a real positive number 
and denote $\fq$ a $*$-Hopf algebra generated by $a$ and $b$,
subject to relations
$$ba=qab, \quad b^*a=qab^*, \quad bb^*=b^*b,\quad a^*a+q^2b^*b=1,\quad aa^*+bb^*=1,\eqno(11a)$$
equipped with a  coproduct
$$\Delta a=a\ot a-qb\ot b^*, \quad \Delta b=b\ot a^* +a\ot b,\eqno(11b)$$
a counit
$\e(a)=1$, $\e(b)=0$ and 
an antipode 
$$Sa=a^*, \quad Sa^*=a,\quad Sb=-qb, \quad Sb^*=-q^{-1}b^*.\eqno(11c)$$
The algebra $\fq$ is thus well defined but it is perhaps 
 useful to comment its name.  As everywhere in this paper, the symbol
 Fun$_q$($M$) indicates
the deformation of the algebra of certain class of functions on the
ordinary manifold $M$.  If the manifold $M$ is the
Lie group then the typical functions in this class are the
matrix elements of the finite-dimensional representations of this group (cf.
 [Levendorskii $\&$ Soibelman]).

The $*$-Hopf algebra $\uq$ is generated by elements  $e$ and (invertible self-adjoint) $k$, 
subject to relations
$$ek=qke,\quad k^2-k^{-2}=(q-q^{-1})(e^*e-ee^*), \eqno(12a)$$
equipped with a coproduct
$$\Delta k=k\ot k, \quad \Delta e=e\ot k+k^{-1}\ot e,\eqno(12b)$$
a counit $\e(k)=1$, $\e(e)=0$ and an antipode
$$Se=-q^{-1}e, \quad  Se^*=- qe^*,\quad Sk=k^{-1}.\eqno(12c)$$
The (non-degenerate) duality pairing between $\uq$ and $\fq$ is given
by the two-dimensional representation of $\uq$, i.e.
$$<k,a>=q^{\jp},\quad  <k,a^*>=q^{-\jp}, \quad <e,-qb^*>=<e^*,b>=1\eqno(12d)$$
with all other couples of generators pairing to $0$.  It is easy to verify
that the star structures on $\fq$ and $\uq$ are compatible in the sense of  Eq. (5).

\vskip1pc

\noindent The Podle\'s sphere is the algebra $\fp$ viewed as the subalgebra of $\fq$
of right invariant elements with respect to the action of the self-adjoint group-like elements
$k$ and $k^{-1}$ (cf. (10)). It is generated by 
$$B=ab,\quad B^*=b^*a^*, \quad A=bb^*,$$
obeying the following relations
$$AB=q^2BA,\quad AB^*=q^{-2}B^*A, \quad BB^*=q^{-2}A(1-A), \quad  B^*B=A(1-q^2A).$$

The action of the $q$-Lorentz group  $D(\uq)$ on $\fp\subset \fq$ is described by  the
formulae  ($7ab$). We obtain explicitely

$$k\tr B=q^{-1}B,   \quad k\tr B^*=qB^*, 
\quad k\tr A=A,$$ 
$$k^{-1}\tr B=qB,\quad  k^{-1}\tr B^*=q^{-1}B^*, \quad k^{-1}\tr A=A,$$
$$e\tr B=0,\quad e\tr B^*=q^{-\jp}-(q^{3\over 2}+q^{-\jp})A,\quad e\tr A=q^{\jp}B,$$
$$e^*\tr B= -q^{-{3\over 2}}+(q^{\jp}+q^{-{3\over 2}})A,\quad e^*\tr B^*=0,
\quad e^*\tr A=-q^{-\jp}B^*,\eqno(13a)$$
$$a\tr B=q^{-1}B+(q-q^{-1})BA,  \quad a\tr B^*=q^{-1}B^*+(q-q^{-1})AB^*,$$
$$ a\tr A=q^{-2}A +(1-q^{-2})A^2,$$
$$a^*\tr B=qB+(q-q^3)AB,  \quad a^*\tr B^*=qB^*+(q-q^{3})B^*A,$$
$$ a^*\tr A=q^{2}A +(q^2-q^{4})A^2,$$
$$b\tr B=(q^2-1)B^2,  \quad b\tr B^*= (1-q^2)A^2,\quad  b \tr A= (q^3-q)BA,$$
$$ b^*\tr B =(q-q^{-1})A^2, \quad b^*\tr B^*=-(q-q^{-1})B^{*2}, \quad b^*\tr A=(1-q^{2})AB^*.\eqno(13b)$$

 We note, that the notion of  $*$-structure is crucial for our
  paper because the group $SL(2,\C)$ (in the context of
  the conformal transformations acting on the Riemann
  sphere) is viewed as the \underline{real} group. It is this fact
    which is the starting point of our strategy to deforme the projective formula
   (1), since the real  group $SL(2,\C)$ is the Poisson-Lie Drinfeld double of the
   group $SU(2)$.
  The concept of reality in the deformed Hopf picture  
  is encoded in the $*$-structure. Thus we need a star $*$ on our quantum double
  $D(\uq)=SL_q(2,\C)$. It is in fact given by the formula
  (6) uniquely in terms of the standard stars on $\uq$ and $\fq$
  (see [Majid, D\c abrowski et al.]). 
     The  star-compatible  action of the $*$-Hopf algebra $D(\uq)$ on the  $*$- algebra $\fq$ (and on its subalgebra $\fp$) 
      is the $q$-deformed version of the statement
  that the real group $SL(2,\C)$ acts on the real algebra Fun$(SU(2))$ and on its subalgebra
  Fun$(S^2)$.

\section{The limit $q\to 1$}
In this section, we want to show that  the action ($7ab$) of the quantum double
$D(\uq)$ on $\fp$ described explicitely by the formulae ($13ab$)  gives in the limit $q\to 1$
the same result as the action of the group $SL(2,\C)$ on Fun$(S^2)$ induced by the 
projective formula  (1).    First of all, 
the limit $q\to 1$ of $\fp$ gives the commutative algebra of complex functions on the sphere $S^2$, generated by 
$$\B={z\over z\bar z+1},\quad \B^*={\bar z\over z\bar z+1} , \quad \A={1\over z\bar z+1},$$
where $z$ is the standard complex coordinate on the Riemann sphere given by the
stereographic projection.

\noindent The subgroup $SU(2)$ of $SL(2,\C)$  acts on $S^2$ via formula (1)
$$z\to {\alpha z+\beta \over -\bar \beta z+\bar \alpha}, \quad \bz\to {\bar\alpha \bz+\bar\beta \over -\beta \bz+\alpha}.$$
Its Lie algebra  $Lie(SU(2))$ therefore acts  on Fun $(S^2)$ via three vector fields $\R_j$, $j=1,2,3$:
$$\R_3= i(z\d-\bar z\dz),\quad \R_1+i\R_2=i(\d+\bz^2\dz),\quad -\R_1+i\R_2=i(z^2\d +\dz).$$
The subgroup $AN$ of $SL(2,C)$ is formed by complex  upper-triangular $2\times 2$-matrices
with real positive numbers on the diagonal. Its action on $S^2$ is obtained from 
the projective formula (1) for the following choice of parameters: $\gamma=0$, Im$\alpha=0$, 
Re$\alpha>0$
and $\beta$ an arbitrary complex number. Thus
$$z\to \alpha(\alpha z+\beta ), \quad \bar z\to   \alpha(\alpha\bz +\bar \beta).$$
The  Lie algebra $Lie(AN)$ therefore acts  on Fun$(S^2)$ via three vector fields $\T_j$, $j=0,1,2$:
$$\T_0= z\d+\bar z\dz,\quad \T_2+i\T_1=-2\dz ,\quad -\T_2+i\T_1= 2\d.$$
It is now straightforward to calculate
$$\R_3\B=i\B,\quad \R_3\B^*=-i\B^*,\quad \R_3\A=0.$$
$$(\R_1+i\R_2)\B=i(2\A-1),\quad (\R_1+i\R_2)\B^*=0, \quad (\R_1+i\R_2)\A=-i\B^*,$$
$$(-\R_1+i\R_2)\B=0,\quad (-\R_1+i\R_2)\B^*=i(2\A-1), \quad (-\R_1+i\R_2)\A=-i\B,\eqno(14a)$$
$$\T_0\B=\B(2\A-1),\quad \T_0\B^*=\B^*(2\A-1),\quad \T_0\A=2\A(\A-1),$$
$$ (\T_2+i\T_1)\B= 2\B^2,\quad (\T_2+i\T_1)\B^*=-2\A^2,\quad (\T_2+i\T_1)\A=2\A\B, $$
$$ (-\T_2+i\T_1)\B= 2\A^2,\quad (-\T_2+i\T_1)\B^*=-2\B^{*2},\quad (-\T_2+i\T_1)\A=-2\A\B^*.\eqno(14b) $$
We recall, that the formulae ($14ab$) describe the infinitesimal projective action (1)
of the  Lie Algebra  $Lie(SL(2,\C))$  on  Fun$(S^2)$. We wish to show that they
can be obtained from the formulae ($13ab$) in the limit $q\to 1$.

In the limit $q\to 1$, the Hopf algebra $\uq$ reduces to the envelopping algebra
of $Lie(SU(2))$. Upon the standard identification 
$$-ie=-R_1+iR_2, \quad ie^*= R_1+iR_2, \quad k=q^{iR_3}, \quad k^{-1}=q^{-iR_3},$$
we indeed obtain  in the limit the standard definition of the $U(su(2))$ (viewed
as the Hopf algebra) from the defining relations  ($12abc$)  of $\uq$. In particular,
the commutations relations ($12a$) gives  in the limit $[R_j,R_k]=\epsilon_{jkl}R_l$.
(Note that $R_j^*=-R_j$.)
In the limit $q\to 1$,  the action ($13a$) of $\uq$  thus gives
$$(R_3 \tr B)_{q\to 1}=lim_{q\to 1}{k-1\over {\rm i ln}q}\tr B= iB,\quad (R_3 \tr B^*)_{q\to 1}=- iB^*,\quad 
(R_3 \tr A)_{q\to 1}=0.$$
$$(( R_1+iR_2)\tr B)_{q\to 1}=\lim_{q\to 1}(ie^*\tr B)= i(2A-1),\quad ((R_1+iR_2)\tr B^*)_{q\to 1} =0,$$
$$((-R_1+iR_2)\tr B)_{q\to 1}=0,\quad ((-R_1+iR_2)\tr B^*)_{q\to 1}=\lim_{q\to 1}(-ie\tr B^*)=
i(2A-1), $$
$$  \quad ((R_1+iR_2)\tr A)_{q\to 1}=-iB^*, \quad ((-R_1+iR_2)\tr A)_{q\to 1}=-iB.\eqno(15a)$$
Comparing ($15a$) with ($14a$), we immediately observe that the $q\to 1$
limit of the $\uq$ action on the Podle\'s sphere indeed coincide with the
$Lie(SU(2))$ action induced by the projective formula.

Now we turn our attention to the $q$-deformation of the  action of  $Lie(AN)$. 
We define the  following elements of $\fq$:
  $$\quad ^qT_0 ={a-a^*\over 2({\rm ln}q)},\quad 
i^qT_1+^qT_2={b\over ({\rm ln}q)},\quad i^qT_1-^qT_2={b^*\over ({\rm ln}q)} \eqno(16a)$$
  and calculate 
  $$\lim_{q\to 1} (^qT_0 \tr B)= B(2A-1),\quad  \lim_{q\to 1} (^qT_0\tr B^*)=B^*(2A-1),\quad 
\lim_{q\to 1} (^qT_0\tr  A)=2A(A-1).$$
$$\lim_{q\to 1}( i^qT_1+^qT_2)\tr B=
 2B^2,\quad \lim_{q\to 1}(i^qT_1+^qT_2)\tr B^* =-2A^2,$$
$$\lim_{q\to 1}((i^qT_1-^qT_2)\tr B
 =2A^2,\quad \lim_{q\to 1}(i^qT_1-^qT_2)\tr B^*=-2B^{*2}, $$
$$  \quad \lim_{q\to 1}(i^qT_1+^qT_2)\tr A=2AB, \quad \lim_{q\to 1}(i^qT_1-^qT_2)\tr A=-2AB^*.\eqno(15b)$$
Comparing ($15b$) with ($14b$), we immediately observe that the $q\to 1$
limit of the $\fq$ action on the Podle\'s sphere indeed gives  the
$Lie(AN)$ action induced by the projective formula (1).

 The reader may find somewhat mysterious why the $q\to 1$
limit  of $\fq$  contains $Lie(AN)$ generators. The explanation of this fact 
resides in the famous Drinfeld duality principle which states that  there is a natural 
 identification
of Hopf algebras Fun$_q(G)$ and $U_q(\G^*)$.  
  Here $G$ is a
Poisson-Lie group  and $\G^*$
is the Lie algebra of its dual Poisson-Lie group $G^*$.  
Let us  indicate (a rigorous proof would require to give meaning
to non-polynomial  functions  appearing in (16b)) why the 
   Drinfeld duality takes place in 
    the case $G=SU(2)$ and $G^*=AN$.   The Lie algebra $Lie(AN)$ is 
generated
by three generators $T_j$, $j=0,1,2$ , $T_j^*=-T_j$ obeying the following commutation 
relations:
$$[T_0,T_1]=-T_1, \quad [T_0,T_2]=-T_2, \quad [T_1,T_2]=0.\eqno(17)$$
 We set 
$$a=q^{T_0}\sqrt{1+q^2({\rm ln}q)^2(T_1^2+T_2^2)},
\quad a^*=\sqrt{1+q^2({\rm ln}q)^2(T_1^2+T_2^2)}q^{-T_0},$$
$$b=({\rm ln}q)(iT_1+T_2), \quad b^*=({\rm ln}q)(iT_1-T_2).\eqno(16b)$$
Then it is not difficult to check two things:
 1) the formulae (16b) and (17) imply the defining commutation 
relations ($11a$) 
 of the Hopf algebra $\fq$;   2)   it holds $\lim_{q\to 1}(^qT_j )=T_j$.

\vskip1pc 
\noindent {\bf Remark 1}:
 Note that 
this explicit relation (16b) between $U_q(Lie(AN))$ and
$\fq$   {\it degenerates} when $q\to 1$. This fact was  important for 
establishing the limit $q\to 1$ of the Hopf adjoint action of $\fq$ 
on 
$\fp\subset \fq$.
Indeed, it appears superficially
that in the $q\to 1$ limit, the algebra $\fq$ becomes commutative 
and
the adjoint action trivial. This observation is too naive, however,
 and 
the explanation
of the paradox resides in the degeneration of the relation (16b)
 between
 the sets of generators
$T_j$ and $a,a^*,b,b^*$ in the limit $q\to 1$.

\vskip1pc
\noindent {\bf Remark 2}: We have established 
 the correct   $q\to 1$ limit of  the quantum double action ($7ab$) by performing the
detailed calculations with the generators, relations etc.  However, it is also possible to  establish  it
on the conceptual level.       First of all, the experts
in Poisson-Lie groups and Hopf algebras know  that the $q\to 1$ 
of the adjoint action ($7b$) of $U_q(\G^*)$ on itself is indeed 
 the dressing transformation of the Poisson-Lie group
 $G$ by its  dual Poisson-Lie group $G^*$ (here $G=SU(2)$ and $G^*=AN$).
    The reader can find the detailed proof of this fact in 
the paper of [Korogodsky] . The conceptual proof of the correct $q\to 1$ limit
    of the formula ($7a$) is even simpler. Indeed:
 
    The standard
left action of the envelopping algebra 
$U(su(2))$  on 
Fun($SU(2)$) is given 
by left  derivations, i.e. if  $X$ is an 
element  of $su(2)$ and 
$h(g)$ is in Fun($SU(2)$) then we have
$$(X\tr h)(g)={d\over dt}h(e^{-tX}g)
\vert_{t=0}.\eqno(18)$$
Recall the coproduct and the counit
of  the Hopf algebra structure of the 
non-deformed Fun($SU(2)$):
$$(\Delta h)(g_1,g_2)=h(g_1g_2), \quad
\varepsilon (h)=h(e),$$
where $e$ is the group unit. Recall also that
 $S(X)=-X$ for $X\in su(2)
\subset U(su(2))$. Finally note the standard 
formula for the pairing $\la.,.\ra$
between $X\in su(2)$ and $h\in$Fun($SU(2)$):
$$-\la X, h\ra=\varepsilon(X\tr h).$$
Putting all these pieces of information together, we
 see that (18) can be written as 
$$(X\triangleright h)(g)={d\over dt}
h(e^{-tX}g)\vert_{t=0}=
 \la S^{-1}(X),h'\ra h''(g).$$
This is  indeed the formula ($7a$) for $H=U(su(2))$ 
and $\tilde H=$Fun($SU(2)$).
In this way, we have verified that
the  action ($7a$) of $\uq$ on $\fq$ (and, consequently on $\fp\subset \fq$ )has the
correct  $q\to 1$ limit, because it is well-known that the left action of $SU(2)$
    on $S^2=SU(2)/U(1)$ is induced by the projective
   formula (1) for    
$\bar \alpha=\delta, \bar \gamma=-\beta$.

\section{Conclusions and outlook}
  We have 
constructed  the  $q$-Lorentz group extension of the natural
action of $\uq$ on the Podle\'s sphere and shown that it  can be 
 naturally interpreted 
as the $q$-deformation of the projective formula 
$z\to {\alpha z+\beta\over \gamma z+\delta}$ describing
the global conformal transformation of the Riemann sphere. 
Our results are rather  mathematical in nature but we believe
that they can be used mainly in mathematical physics e.g. 
  in further studies  of braided field theories 
 (cf. [Oeckl])
and also in studies of $q$-differential operators 
(cf. the $q$-Dirac operator
by [D\c abrowski $\&$ Sitarz])
 on the Podle\'s sphere. Indeed, our studies suggest  to investigate the 
symmetry properties
of those objects not only from the point of view of the action of
 the 
 $\uq$ 
quantum group but also from the point of view of the action of its  
quantum double.

\vskip2pc
\noindent {\bf Acknowledgement}: I thank to R. Oeckl for
 discussions and to A. Sitarz for providing me the draft of
his paper [D\c abrowski et al] prior to publication.

\vskip1pc
\noindent {\bf Note added}: After having posted to the arch-ive  the second version o
f this paper, S. Woronowicz has attracted my attention to his joint paper with W. Pusz 
in which they induced the representations of the $q$-Lorentz group  from the characters
of its parabolic subgroup. Among the representations constructed in this way there is also
 one that corresponds to the action of the $q$-Lorentz group on the Podle\'s sphere.

 \newpage
\noindent {\large  \bf References}:
\vskip2pc

\noindent [Drinfeld] V. G. Drinfeld, {\it Quantum groups} in Proc. 
ICM, MSRI, Berkeley, 1986;

\noindent [D\c abrowski $\&$ Sitarz] L. D\c abrowski and A. Sitarz,
{\it Dirac operator on the standard Podle\'s quantum sphere}, 
Noncommutative geometry and quantum groups (Warsaw, 2001),
49-58, Banach Center Publ., 61, Polish Acad. Sci., Warsaw, 2003, 
math.QA/0209048,

\noindent [D\c abrowski et al] L. D\c abrowski, G. Landi , A. Sitarz, W. van Suijlekom and J. Varilly,
{\it The Dirac operator on $SU_q(2)$}, to appear

 \noindent[Kassel] C. Kassel, {\it Quantum groups}, Springer-Verlag, New York, 1995;

\noindent [Klim\v c\'\i k] C. Klim\v c\'\i k, {\it Quasitriangular 
WZW model}, Sec. 4.1,   Rev. Math. Phys. {\bf 16}, (2004) 679 - 808,  hep-th/0103118

 \noindent [Korogodsky] L. Korogodsky, {\it Complementary series 
representations
and quantum orbit method}, q-alg/9708026;

\noindent[Levendorskii $\&$ Soibelman] S. Levendorskii and Y. Soibelman,
{\it Algebras of functions on compact quantum groups, Schubert cells and quantum tori},
 Commun. Math. Phys. {\bf 139}  141 - 169 (2001);

\noindent [Majid]  S. Majid, {\it Foundations of quantum group 
theory}, Cambridge University
Press , Cambridge, 1995;

\noindent [Oeckl] R. Oeckl, {\it Braided quantum  field theory}, 
Commun. Math. Phys. {\bf 217}  451 - 473 (2001);

\noindent [Podle\'s] P. Podle\'s, {\it Quantum spheres}, Lett. 
Math. Phys. {\bf 14}  521 - 531 (1987);

\noindent [Podle\'s $\&$ Woronowicz] P. Podle\'s and S. Woronowicz,
{\it Quantum deformation of Lorentz group}, Commun. Math. Phys. 
{\bf 130}  381 - 431 (1990);
 
\noindent[Pusz $\&$  Woronowicz] W. Pusz and S. Woronowicz,
 {\it  Representations of Quantum Lorentz Group on Gelfand spaces}, 
 Rev. Math. Phys. {\bf 12}  1551-1625 (2000)

\noindent [Semenov-Tian-Shansky] M. A. Semenov-Tian-Shansky, 
{\it Dressing transformations and Poisson group actions}, Publ. 
RIMS, Kyoto University {\bf 21}  1237 - 1260 (1985).

\end{document}